\documentclass{elsart}

\usepackage{graphicx}

\newcommand{\beq}{\begin{equation}}
\newcommand{\eeq}{\end{equation}}

\begin{document}

\begin{frontmatter}

\title{NMR spectroscopy of hydrogen deuteride and 
magnetic moments of deuteron and triton}

\author[litmo]{Yurii I. Neronov}
\and
\author[vniim,mpq]{Savely G. Karshenboim}
\ead{sek@mpq.mpg.de}
\address[litmo]{St. Petersburg State Institute of Fine Mechanics 
and Optics, St. Petersburg, Sablinskaya ul. 14, St.Petersburg 197101, Russia}
\address[vniim]{D. I. Mendeleev Institute for Metrology (VNIIM),
 St. Petersburg 198005, Russia}
\address[mpq]{Max-Planck-Institut f\"ur Quantenoptik, 85748 Garching, Germany}

\begin{abstract}
Magnetic moments of free and bound deuteron and triton are 
considered and new results for their magnetic moments 
(in units of that of the proton) and their $g$ factors are 
presented. We report on a measurement with medium-pressure hydrogen 
deuteride (HD) at 10~atm, which is to be compared with the 
previous measurement done at 100~atm. We confirm that the high 
pressure used in former experiments caused no systematic effects at a level of 10~ppb. 
We also reexamined a theoretical uncertainty related to screening 
effects in HD and HT molecules and found that previously it was underestimated. 
The medium-pressure result obtained for the free deuteron 
$\mu_d/\mu_p=0.307\,012\,206\,5(28)$ with a fractional uncertainty 
of $9.1 \cdot10^{-9}$ is free of systematic effects related to 
former high-pressure experiments. The reevaluated result for 
triton is $\mu_t/\mu_p=1.066\,639\,908(10)$ with a fractional 
uncertainty of $9.3 \cdot10^{-9}$.
\end{abstract}


\end{frontmatter}

Magnetic moments of light nuclei have been accurately studied for 
quite a while. 
Their studies play an important role since they contribute to several 
high-precision measurements. First, to determine a magnetic moment 
of a given object (particle, nucleus etc.) via a measurement of the 
spin precession frequency, one has to use some known magnetic moment and 
light nuclei supply us with such data (see e.g. the compilation 
\cite{adndt,green,fire}). Second, to determine the fine structure 
constant $\alpha$, one can measure a gyromagnetic ratio of some nucleus 
with a value of the magnetic moment known in units of the nuclear magneton 
or Bohr magneton (see e.g. original results on the gyromagnetic 
ratio of proton \cite{gyrop}, helion \cite{gyrohe} and 
reviews \cite{codata-,codata} for detail). Recently magnetic 
moments of light nuclei ($A\leq 3$) were in part revisited 
\cite{ki_plb} to take into account higher-order binding effects and some 
new experimental data related to the nuclear magneton value. Here we 
continue our reexamination of existing data and present a new 
measurement of the deuteron magnetic moment.

Studies of the deuteron magnetic moment are of particular interest 
since this is the simplest nucleus which consists of more than one nucleon. 
The most accurate results on the magnetic moment of the 
proton \cite{wink} and  
deuteron \cite{phil84} adopted in \cite{codata} and recommended 
by CODATA are based on MIT experiments with hydrogen (or deuterium) 
atoms with a maser. The energy levels studied were split by the 
magnetic field. 
Because of the significance of the MIT results, an independent determination 
of the ratio of proton-to-deuteron magnetic moments is 
of crucial importance. The magnetic moment of deuteron (proton) can 
be determined in one of the following ways:
\begin{itemize}
\item from the splitting of the $1s$ sublevels in deuterium 
(hydrogen) as it was done at MIT with an 
uncertainty at a level of a part in $10^8$ \cite{wink,phil84};
\item by measuring a spin precession frequency of a free deuteron 
(proton) as it was done for a proton with an uncertainty not better than 
a part in $10^7$ for a number of cases (mainly to calibrate the 
magnetic field while measuring magnetic moments of other objects);
\item by studying a spin precession frequency related to proton 
and deuteron for a HD molecule as we discuss in detail here.
\end{itemize}

The magnetic resonance spectroscopy of H$_2$, D$_2$ and HD to 
determine the magnetic moments of proton and deuteron as well as the 
deuteron electric quadrupole moment was first applied in \cite{first} 
about sixty years ago. Later, important advantages of the application 
of a HD molecule for determination of $\mu_p/\mu_d$ were pointed 
out by Ramsey \cite{ramsey_hd}. A description of the NMR spectrum of HD 
can be found in detail in e.g. \cite{karplus}. The results of 
early experiments performed about fifty years ago are discussed 
in \cite{beam}. Since that time the accuracy has been substantially 
improved and is now claimed to be at a level of a few ppb. 
However, the most accurate molecular experiments on 
HD \cite{neronov_old,neronov_he,neronov_d} and 
HT \cite{neronov_t} have not been considered 
in Ref. \cite{ki_plb} following the statement of \cite{codata} 
on their unclear uncertainty, while they were used in other 
compilations \cite{adndt,green,fire,codata-}. The earlier 
molecular experiments (see e.g. \cite{early_hd}) are less accurate 
and are not competitive.

We believe that potentially the quoted experiments can provide us 
with the most accurate data and intend to clarify their uncertainty. 
Here we reconsider theoretical and experimental uncertainties 
related to the NMR spectroscopy of molecular hydrogen
and its isotopic modifications and 
report on the new experimental result for hydrogen deuteride. A crucial 
level of accuracy is a part in $10^8$ which is needed to verify the
results of the MIT experiments \cite{wink,phil84} and our goal 
is to reach at least this level.

In NMR experiments on molecular hydrogen and its isotopes 
performed in 1975--1989 \cite{neronov_old,neronov_he,neronov_d,neronov_t} 
several quantities were determined with an extremely high accuracy: 
$\mu_p({\rm H}_2)/\mu_p({\rm HD})$, $\mu_d({\rm HD})/\mu_d({\rm D}_2)$, 
$\mu_d({\rm HD})/\mu_p({\rm HD})$, $\mu_t({\rm HT})/\mu_p({\rm HT})$. 
The claimed uncertainty was at a level of a few parts in $10^9$. 
A signal related to $\mu_p({\rm H}_2)$ in the tritium experiment 
\cite{neronov_t} was also observed, however, the value of 
$\mu_p({\rm H}_2)/\mu_p({\rm HT})$ was not determined since the pure 
hydrogen signal was subtracted and the data are not available any
longer.

The experiments \cite{neronov_he,neronov_d,neronov_t} were performed 
at a high pressure of approximately $10^7$ Pa (100 atm) and often with 
a sample rotation. The purpose of this paper is to check whether 
rotation and high pressure affect the results. The experiment reported 
here is performed on HD and D$_2$ molecules at a pressure of 
$10^6$~Pa (10~atm) without a sample rotation. We also 
reexamine a theoretical uncertainty appearing due to the interpretation 
of the proton-to-deuteron ratio at molecular HD in terms of those 
for free nuclei.

The most important results of the high-pressure 
experiments \cite{neronov_old,neronov_he,neronov_d,neronov_t} 
are summarized in Table~\ref{Tab_high}, which contains 
all reliable data. The ratio of nuclear magnetic moments is easily 
related to measured NMR frequencies
\begin{equation}\label{ratio}
\frac{\mu_A}{\mu_B} = \frac{f_A}{f_B} \,\frac{I_A}{I_B}\;, 
\end{equation}
where $\mu$ is the magnetic moment, $I$ is the nuclear spin and $f$ is the 
NMR frequency. A value of the magnetic moment there is a bound one
\begin{equation}\label{free}
\mu({\rm bound}) = \mu({\rm free}) \cdot \Bigl(1-\sigma\Bigr)\;.
\end{equation}
The uncertainty in the calculation of the screening constant $\sigma$ 
are discussed separately below. A value of the NMR 
frequency $f$ is the unweighted average over the multiplet: 
a triplet for proton in HD, a doublet for deuteron in HD, triton and 
proton in HT, a singlet for H$_2$ and D$_2$. The separation between 
the lines within a multiplet is determined by the constant $J$ related 
to the nuclear spin-spin interaction (see e.g. \cite{karplus}).

The measurements were taken at room temperature. The influence of the  
rotational and vibrational structure on the screening constants 
was studied at room temperature in \cite{neronov_he}.

\begin{table}[hbtp]
\caption{Results of the high pressure experiments. Unless otherwise
stated a sample was rotated. The characteristic 
rotation frequency $\omega_{\rm rot}=5-10$ Hz and the rotation axis 
was perpendicular to the magnetic field. \label{Tab_high}}
\begin{center}
\begin{tabular}{llllll}
\hline
Quantity & Value & Magnetic & Pressure,& Comment & Ref. \\
 &  &  field, T & atm & & \\
\hline
$\mu_p({\rm HD})/\mu_p({\rm H}_2)$ & $1-35.9(2)\times 10^{-9}$ 
& 1.52 & 130  & & \protect\cite{neronov_old} \\
$\mu_d({\rm D}_2)/\mu_d({\rm HD})$ & $1-42(2)\times 10^{-9}$ 
& 1.52 & 130  & & \protect\cite{neronov_old} \\
$J({\rm HD})$ & 43.115(12) Hz & 1.52 & 130  & 
& \protect\cite{neronov_old} \\
$\mu_t({\rm HT})/\mu_p({\rm HT})$ & 1.066\,639\,887(3)& 1.49 
& 100  & No rotation & \protect\cite{neronov_t} \\
$J({\rm HT})$ & 299.3(2) Hz & 1.49 & 100  & 
& \protect\cite{neronov_t} \\
$\mu_{h}(^3{\rm He})/\mu_p({\rm H}_2)$ & 0.761\,786\,637(13) 
& 1.49 & 60  & & \protect\cite{neronov_he} \\
$\mu_{h}(^3{\rm He})/\mu_p({\rm HD})$ & 0.761\,786\,661(5) 
& 1.49 & 78  & & \protect\cite{neronov_he} \\
$\mu_{h}(^3{\rm He})/\mu_d({\rm HD})$ &   
2.481\,291\,l25(3) & 1.49 & 78  & & \protect\cite{neronov_he} \\
$\mu_p({\rm HD})/\mu_d({\rm HD})$ & 
3.257\,199\,655(45) & 1.49 & 78  & Mixture with He 
& \protect\cite{neronov_he} \\
$\mu_p({\rm HD})/\mu_d({\rm HD})$ & 
3.257\,199\,515(40) & 1.41 & 130  & & \protect\cite{neronov_d}\\
$\mu_p({\rm HD})/\mu_d({\rm HD})$ & 
3.257\,199\,496(20) & 4.70 & 100  & No rotation 
& \protect\cite{neronov_d} \\
$\mu_p({\rm HD})/\mu_d({\rm HD})$ & 
3.257\,199\,514(4) & 4.70 & 100  & & \protect\cite{neronov_d}\\
\hline
\end{tabular}
\end{center}
\end{table}

During the experiment described in \cite{neronov_old} the result on 
${\mu_d}/{\mu_p}$ was also derived, however, later it was found to be 
contradicting to MIT data for a free proton and 
deuteron \cite{wink,phil84} (see \cite{codata-,codata} for detail) 
\begin{equation}\label{mupdmaser}
\frac{\mu_d}{\mu_p}=0.307\,012\,207(5)\;,
\end{equation}
which has a fractional uncertainty of $1.5\cdot10^{-8}$.
A search for systematic effects led to understanding that commonly used 
techniques of separated inductance coils for proton and deuteron resonance 
provide the excitation of proton and deuteron transitions in slightly 
different areas and due to an inhomogeneity of the magnetic 
field Eq.~(\ref{free}) is not valid any longer and the ratio of two 
NMR frequencies is slightly different from the ratio of the corresponding 
magnetic moments. The systematic effect was explained in the last 
paper of series \cite{neronov_d}, however, a technique with a use of the 
same inductance for both resonances was already applied in the experiments 
with triton \cite{neronov_t} and helion \cite{neronov_he}. However, 
even after fixing this systematic source, some other questions remain to be 
clarified. In particular, one has to check effects due to a sample
rotation, high pressure and uncertainty of theory. 

Effects of the rotation were studied theoretically by 
Rabi, Ramsey and Schwinger in \cite{rotation} (see also \cite{beam}) 
who found that a particle in the rotated frame feels an 
effective magnetic field different from that in the rest frame.
The value of the difference depends on a gyromagnetic ratio of the 
particle and thus different particles see a somewhat different 
effective field, so that Eq.~(\ref{ratio}) is not valid any longer. 
The ratio of the spin precession frequencies is different from that 
containing the magnetic moments only. The effective correction to 
the field is additive and proportional to the rotation frequency, 
which in the former high-pressure experiments 
\cite{neronov_old,neronov_he,neronov_d} was perpendicular to the 
magnetic field. Hence the linear effect in 
$\omega_{\rm rot}/\omega_{\rm NMR}$ vanishes as long as the 
rotation axis is perpendicular to the magnetic field and only 
quadratic effects, which are negligible, are present. 
However, a linear effect
in $\omega_{\rm rot}/\omega_{\rm NMR}$ of the reduced value
could appear if the field and the axis are not 
exactly perpendicular. The rotation could also cause some mechanical 
and chemical problems. One of the quoted high-pressure experiments 
\cite{neronov_d} observed no shift of $\mu_p({\rm HD})/\mu_d({\rm HD})$ due to the rotation at a level 
of $6\cdot10^{-9}$. In our experiment we also do not apply 
any rotation.

To check the pressure dependence, we performed a medium-pressure 
experiment at 10~atm results of which are reported here. This is approximately 
tenfold reduction compared to previous high-pressure experiments 
done at a pressure from 60 to 130 atm, with the most accurate 
results obtained at the highest applied pressure. A higher 
pressure as well as rotation served well to reduce a line width of 
NMR signals and in former high-pressure experiments there was 
no overlap between lines within multiplets (a triplet for proton 
and a doublet for deuteron in HD with a deuteron singlet from D$_2$). 
In the present experiment the lines within multiplets are overlapping. 

Let us briefly describe our experiment. We used a commercial 
NMR spectrometer Bruker CXP 300 with a superconductive magnet 
producing the magnetic field of 7.05~T. The sample consisted 
of an external tube (\#508-UP by Norell, Inc.) and an internal 
capillary. The internal capillary was filled with gas of HD and 
D$_2$ at a pressure of 10~atm. The signal of the gas sample is 
relatively weak and we need some time to collect reliable data. 
In our experiment the measurement time was one hour. To simplify 
a calibration procedure, we 
filled the external tube with acetone (CH$_3$)$_2$CO and 
deutero-acetone (CD$_3$)$_2$CO.

Acetone provides strong and narrow lines and we controlled 
the magnetic field by fixing the position of the line of deutero-acetone
\begin{equation}
f_d(({\rm CD}_3)_2{\rm CO})= 46\,072\,631.58(1)\;{\rm Hz}
\end{equation}
and measured the acetone signal
\begin{equation}
f_p(({\rm CH}_3)_2{\rm CO})= 300\,135\,552.4(1)\;{\rm Hz}\;.
\end{equation}

\begin{figure}[hbtp]
\begin{center}
\includegraphics[width=0.95\textwidth]{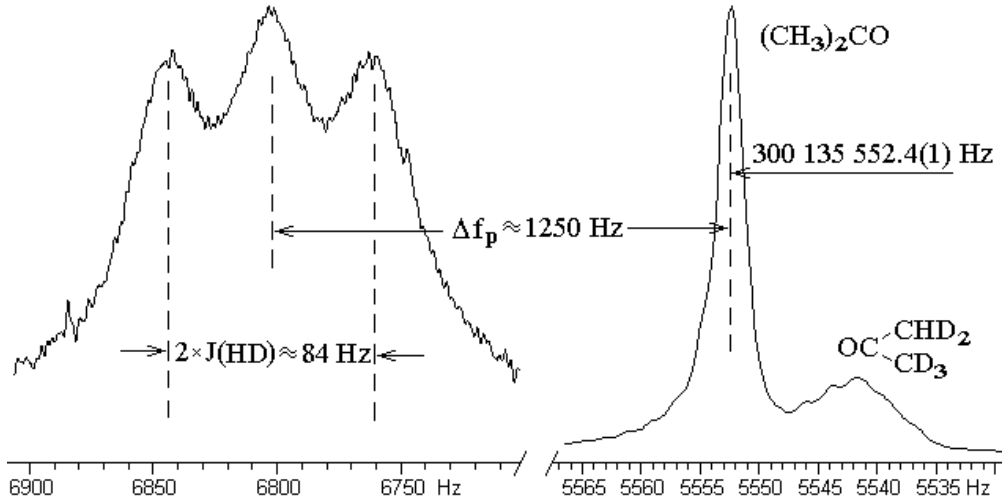}
\end{center}
\caption{The proton NMR spectrum: acetone lines and HD triplet.}
\end{figure}

At the next step we studied the proton triplet lines (see Fig.~1) 
shifted from the acetone line by 
\begin{equation}
\Delta f_p = f_p({\rm HD})- f_p(({\rm CH}_3)_2{\rm CO}) = 1\,250.3(5)\,{\rm Hz}
\end{equation}
with the line separation within the triplet related to the hyperfine 
nuclear spin-spin constant $J({\rm HD})$ and discussed below. 

The other measurement was related to the deuteron lines (see Fig.~2). 
We determined the NMR frequency of the deuteron HD doublet 
shifted from the deutero-acetone line by
\begin{equation}
\Delta f_d = f_d({\rm HD})- f_d(({\rm CD}_3)_2{\rm CO}) = 198.8(4)\,{\rm Hz}\;,
\end{equation}
with the line separation within the doublet determined by 
$J({\rm HD})$, and the deuteron singlet line from D$_2$ 
slightly shifted with respect to the center of the deuteron doublet 
by approximately 2~Hz.

\begin{figure}[hbtp]
\begin{center}
\includegraphics[width=0.95\textwidth]{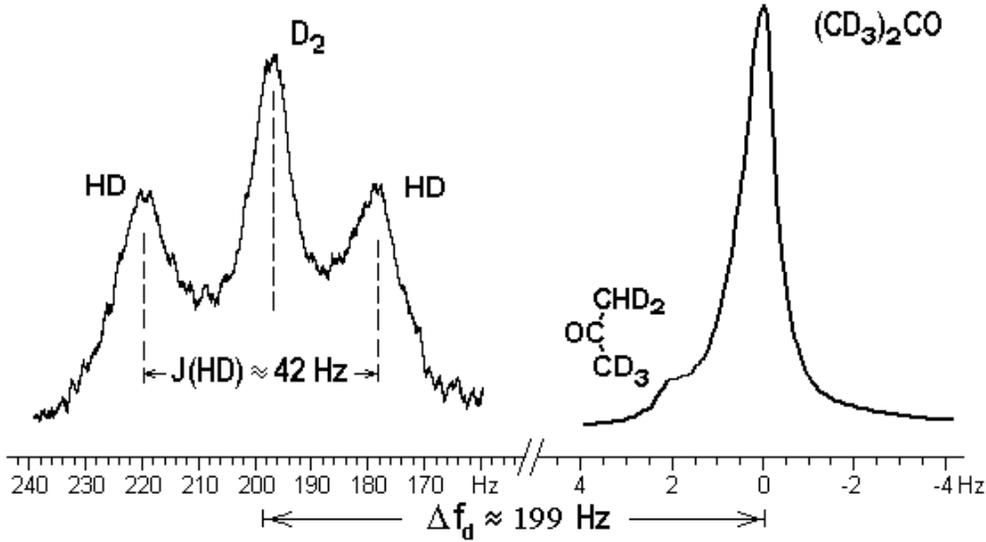}
\end{center}
\caption{The deuteron NMR spectrum: deutero-aceton lines, 
HD doublet and D$_2$ singlet.}
\end{figure}

With these frequencies measured we arrive to a result
\begin{eqnarray}\label{mupmudhd}
\frac{\mu_p({\rm HD})}{\mu_d({\rm HD})}
&=& \frac{1}{2}\;\frac{f_p({\rm(CH}_3{)}_2{\rm CO})+
\Delta f_p}{f_d({\rm(CD}_3{)}_2{\rm CO})+\Delta f_d }\nonumber\\
&=& 3.257\,199\,49(3)
\end{eqnarray}
with a fractional uncertainty of 9 ppb. The uncertainty is mainly 
due to a line fitting procedure. The details will be published 
elsewhere \cite{preparation}. We hope that by improving  
statistics the errors due to the fitting will be reduced.

To check our results, we have performed several comparisons with 
previous results, which are discussed below.

\underline{Acetone.} The acetone ratio ($f_p/f_d$) is found to be 
in  fair agreement with the data previously obtained at a lower magnetic 
field of 1.5~T and 4.7~T with rotation \cite{neronov_a1,neronov_a2}. 
In the present experiment the acetone frequencies are related 
to the position of the maximum of the lines, while for an interpretation 
in terms of magnetic moments one has to separate the dominant proton 
and deuteron modes and  weak contributions related to molecules 
with incomplete substitution of hydrogen for deuterium, 
(CD)$_3$(CHD$_2$)CO, which affect the maximum of stronger lines. 
Such a correction can be different under different conditions 
and the ratio for the quoted proton and deuteron frequencies is not 
exactly the same as the ratio of the proton and deuteron magnetic moments 
in acetone molecules. In this paper we have not considered this 
small correction because the acetone measurements were used for 
the normalization only and the interpretation of the line maximum 
is not important for such a purpose. 

\underline{Hyperfine interaction constant $J$(HD).} The proton 
triplet should consist of three Lorentzian lines separated by the same 
interval as the two lines of the deuteron doublet related to the
HD molecule (see Figs.~1 and~2). The separation is equal to the 
hyperfine constant $J({\rm HD})$ (see e.g. \cite{karplus}). The present results are 
41.8(17)~Hz (from the proton spectrum) and 40.8(15)~Hz which both 
agree with a high-pressure result of 43.115(12)~Hz \cite{neronov_old} 
listed in Table~\ref{Tab_high}.

\underline {Isotopic shift for deuteron.} Simultaneous observation 
of the deuteron singlet related to D$_2$ and the deuteron 
HD doublet provides us with another test (see Fig.~2). The result 
obtained is 
\begin{equation}
\frac{\mu_d({\rm D}_2)}{\mu_d({\rm HD})}= 1-45(7)\times 10^{-9}
\end{equation}
which is less accurate but consistent with the high-pressure 
value of $1-42(2)\times 10^{-9}$ \cite{neronov_old}.

All tests showed  fair agreement with high-pressure results obtained 
applying a sample rotation. The main result of this paper in 
Eq.~(\ref{mupmudhd}) also agrees with the high-pressure values 
listed in Table~\ref{Tab_high}. 
The result (\ref{mupmudhd}) is more reliable than the tests with 
the isotopic effects and the hyperfine structure. 
The result (\ref{mupmudhd}) comes from average frequencies of 
proton triplet and deuteron doublet which are relatively immune 
to line-shape effects. In contrast, differential characteristics 
like the hyperfine separation inside the triplet and the 
doublet or the isotopic shift of the center of gravity of 
the deuteron doublet with respect to the singlet from D$_2$ are 
sensitive to such effects.

Theoretical calculations for shielding and hyperfine constants were 
started by Ramsey \cite{ramsey_th,ramsey_hfs}. Now they are 
substantially improved. We follow the examination in \cite{neronov_t}, 
however, we pay our attention to a proper estimation of the uncertainty 
of the quoted calculation.

The result presented in paper \cite{neronov_t} stated
\begin{eqnarray}
\sigma_t({\rm HT}) - \sigma_p({\rm HT}) &=& 20.4 \times 10^{-9}\;,
\label{s_old_t}\\
\sigma_d({\rm HD}) - \sigma_p({\rm HD}) &=& 15.0 \times 10^{-9}\;.
\label{s_old_d}
\end{eqnarray}
This result was derived as an evaluation of some matrix elements 
with the wave functions related to the Born-Oppenheimer approximation. 
All the perturbative operators to be averaged are of the order of 
$\alpha^2 (m_e/m_p)$ in atomic units. 

The uncertainty of the calculation was not presented explicitly 
in \cite{neronov_t} and the final results for deuteron and triton 
did not include any theoretical uncertainty. However, at least 
two sources of uncertainties were mentioned by the authors and we 
consider them here. Note that any detail in \cite{neronov_t} is 
related to HT, while here we concentrate on HD, and thence 
any contribution to the HD case is multiplied by roughly 0.75 as explained 
in \cite{neronov_t}.
\begin{itemize}
\item The result \cite{neronov_t} is a sum of eight terms. 
However, three of them were estimated as being smaller than 
$4.5\cdot 10^{-10}$ each. An additional term discussed there was estimated 
to be below than $1.5\cdot 10^{-10}$. Since no details on the 
three terms are presented, we consider the missed terms as correlated 
and estimate their sum as $\pm1.5\cdot 10^{-9}$ or approximately 
10\% of the value of $\sigma_d({\rm HD}) - \sigma_p({\rm HD})$. 
\item The wave functions in \cite{neronov_t} are related to 
the binding energy which is approximately 5\% different from a 
real one and due to that authors stated that the uncertainty 
cannot be lower than 5\%. 
\end{itemize}
There may be some other sources due to the method of the 
calculation in \cite{neronov_t} which are beyond consideration in this paper.
We hope to decrease the experimental uncertainty for the ratio 
of the magnetic moments to a few ppb and thus the theoretical 
calculations need to be substantially improved.

Here we estimate the uncertainty of the calculation 
in Ref. \cite{neronov_t} as 15\% and thus the results (\ref{s_old_t}) 
and (\ref{s_old_d}) of \cite{neronov_t} should be corrected to
\begin{eqnarray}
\sigma_t({\rm HT}) - \sigma_p({\rm HT}) &=& 20(3) \times 10^{-9}\;,
\label{sigmath}\\
\sigma_d({\rm HD}) - \sigma_p({\rm HD}) &=& 15(2) \times 10^{-9}\;.
\label{sigmadh}
\end{eqnarray}

Due to the theoretical correction (\ref{sigmadh}), the result 
(\ref{mupmudhd}) leads for a free proton and deuteron:
\begin{eqnarray}\label{mupmudfree}
\frac{\mu_p({\rm free})}{\mu_d({\rm free})} 
&=& 
\frac{\mu_p({\rm HD})}{\mu_d({\rm HD})}\times 
\Bigl(1 - \bigl[\sigma_d({\rm HD}) - \sigma_p({\rm HD})\bigr]\Bigr) 
\nonumber\\
&=& 3.257\,199\,44(3)
\end{eqnarray}
and
\begin{equation}
\frac{\mu_d({\rm free})}{\mu_p({\rm free})} = 0.307\,012\,210(3) 
\end{equation}
with a fractional uncertainty of 9 ppb, which agrees with both 
high-pressure data and the maser result. We summarize results on determination of the deuteron-to-proton ratio in Fig.~3.

\begin{figure}[hbtp]
\begin{center}
\includegraphics[width=0.6\textwidth]{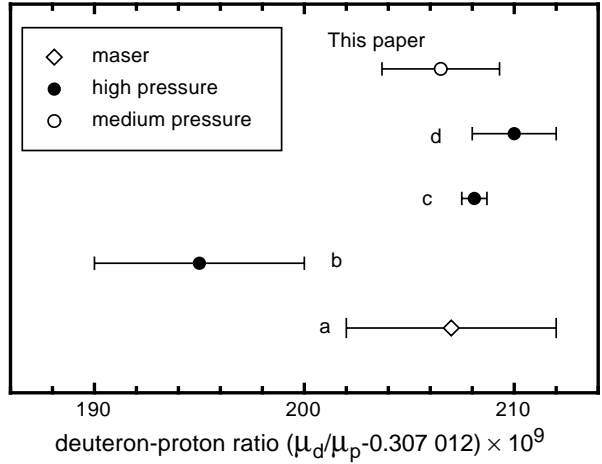}
\end{center}
\caption{The deuteron-to-proton ratio of the magnetic moments: 
the maser result {\em a} is taken from \protect\cite{codata}, 
the result {\em b} is related to an experiment with admixure of helium, 
while the results {\em c} and {\em d} correspond to the measurement 
at a pressure of 100~atm with/without rotation of a sample. The 
high-pressure results have an uncertainty corrected according to 
Eq.~(\protect\ref{sigmath}). 
See Table~1 for references and other detail.}
\end{figure}

Mixing different gaseous substances, one can cancel systematic 
sources due to macroscopic screening effects, perturbations of 
electronic wave functions etc. and to compare a given magnetic 
moment with a probe one. A promising probe magnetic moment is that 
in molecular hydrogen, which is one of very few atomic and molecular 
systems where the binding corrections to the nuclear magnetic moment 
can be in principle calculated {\em ab initio} with a high accuracy. 
In this way it is very important to have more experimental data. 
A study of isotopic and hyperfine effects is an important step 
towards theoretical understanding of the molecular screening 
effects. Meanwhile, the comparison of the proton-to-deuteron 
(and proton-to-triton) ratio of magnetic moments related to 
the same molecule allows accurate relative measurements 
with systematic effects essentially reduced. 
We have demonstrated that the high-pressure data 
obtained for HD molecules are reliable and confirmed them 
within the accuracy of our experiment. We have also carefully 
investigated a theoretical uncertainty which happens to be higher 
than a statistical uncertainty of the former high-pressure experiments \cite{neronov_d,neronov_t}, 
but below the uncertainty of the present medium-pressure experiment. 

\begin{table}[hbtp]
\caption{Final results obtained at medium-pressure. The deuteron 
results are achieved experimentally. The triton results are 
obtained adjusting the uncertainty. \label{T_final}}
\begin{center}
\begin{tabular}{llc}
\hline
Quantity & Value & Fractional uncertainty\\
\hline
$\mu_p({\rm HD})/\mu_d({\rm HD})$ & 3.257\,199\,531(29)      
& $8.8 \cdot10^{-9}$  \\
$\mu_t({\rm HT})/\mu_p({\rm HT})$ & 1.066\,639\,887(10)    
& $8.8 \cdot10^{-9}$  \\
$\mu_p({\rm free})/\mu_d({\rm free})$ &  3.257\,199\,482(30) 
& $9.1 \cdot10^{-9}$ \\
$\mu_d ({\rm free}) /\mu_p({\rm free})$ &  0.307\,012\,206\,5(28) 
& $9.1 \cdot10^{-9}$  \\
$\mu_t({\rm free})/\mu_p({\rm free})$ & 1.066\,639\,908(10)    
& $9.3 \cdot10^{-9}$  \\
$g_d({\rm free}) $ & 0.857\,438\,240(12) & $1.3 \cdot10^{-8}$  \\
$g_t({\rm free}) $ & 2.978\,962\,44(4)   & $1.4 \cdot10^{-8}$ \\
\hline
\end{tabular}
\end{center}
\end{table}

We list in Table~\ref{T_final} the results for deuteron obtained in the 
present paper and the reexamined results for HT. We estimate 
the experimental HT uncertainty by the uncertainty of the present medium-pressure 
experiment since no discrepancy between medium-pressure and high-pressure 
data for HD has been detected. The theoretical uncertainty is taken from (\ref{sigmath}). Even with the reestimated and 
thus enlarged uncertainty the triton result obtained from HT is 
the most accurate one (cf. \cite{duffy,rawi}) and the deuteron result 
has approximately an 1.5 times lower uncertainty than the maser result 
(\ref{mupdmaser}) \cite{wink,phil84,codata}. We consider our 
experimental uncertainty here as rather somewhat overestimated and 
hope to improve medium-pressure measurements. We note, however, 
that presently the theory also substantially contributes to the 
uncertainty for the free magnetic ratio and further theoretical 
progress to obtain more reliable and more accurate data is strongly needed.

The experiment has been performed with the NMR spectrometer at 
the State Research Institute of Highly Pure Biopreparations 
(St.~Petersburg) and the authors are grateful 
to B. P. Nikolaev and Yu. N. Tolparov for their hospitality 
and cooperation. The authors would like to thank Anatoly Barzakh, 
Jim Friar, Peter Mohr and Victor Yakhontov for useful and 
stimulating discussions. The work of SGK was supported in part 
by RFBR under grant 02-02-07027.

\end{document}